\documentclass[conference]{IEEEtran}
\IEEEoverridecommandlockouts
\usepackage{cite}
\usepackage{amsmath,amssymb,amsfonts}
\usepackage{algorithmic}
\usepackage{graphicx}
\usepackage{textcomp}
\usepackage{xcolor}
\usepackage{booktabs}
\usepackage{multirow}
\def\BibTeX{{\rm B\kern-.05em{\sc i\kern-.025em b}\kern-.08em
    T\kern-.1667em\lower.7ex\hbox{E}\kern-.125emX}}

\begin{document}

\title{Grid-Mode-Aware AI Data Center Power Smoothing via Bayesian-Optimized Hybrid Energy Storage System\\
\vspace{-5mm}
}

\author{\IEEEauthorblockN{Sina Mohammadi, Quang Manh Hoang, Marcus Chen I Wada, Van-Hai Bui, Wencong Su}
\IEEEauthorblockA{\textit{Department of Electrical and Computer Engineering} \\
\textit{University of Michigan-Dearborn}\\
Dearborn, USA \\
{sinamo, manhhq, wadamc, vhbui, wencong}@umich.edu}
\vspace{-12mm}
}

\maketitle

\begin{abstract}

AI data centers exhibit highly stochastic, volatile power demands from intensive workloads and synchronized GPU training cycles, which can excite grid natural frequencies and threaten system stability. This paper proposes a real-time, forecasting-independent control framework utilizing a Hybrid Energy Storage System (HESS) that coordinates a Battery Energy Storage System (BESS) and a supercapacitor (SC). Employing a parameterized frequency-decoupling and active modal-damping policy, the framework mitigates critical oscillatory load components while smoothing overall power fluctuations. To bypass the heavy computational burden of online receding-horizon optimization, this work introduces an offline Bayesian Optimization (BO) approach to optimally tune real-time control parameters, minimizing a multi-objective cost function that balances grid compliance penalties and storage degradation. Numerical simulations demonstrate that the proposed method ensures grid stability and strong target envelope adherence. Specifically, it reshapes extreme 75–160 MW raw demand fluctuations to largely fit within a tight 10 MW target band, with only brief transient excursions, while attenuating the critical 20 MW oscillatory component below 1 MW to suppress resonant excitation.

\end{abstract}


\vspace{-5mm}
\section{Introduction}
The rapid growth of artificial intelligence (AI) is increasing both the
number and the size of data centers (DCs). According to recent reports,
more than 11,000 DCs were operating worldwide by early 2024, and their
electricity consumption has become a noticeable part of the total power
demand. In the United States, server electricity use increased from about
30~TWh in 2014 to nearly 100~TWh in 2023, mainly due to the rapid growth
of AI computing infrastructure
\cite{shehabi20242024,valverde2025forced}.

Besides their high energy consumption, AI DCs can also create large and
fast changes in power demand. During AI training, a large number of GPUs
may switch together among computation, communication, and checkpointing
stages, resulting in repeated and highly variable power fluctuations.
These variations can make real-time power balancing and frequency
regulation more difficult. More importantly, periodic or quasi-periodic
load variations can continuously excite a poorly damped grid mode when
their frequency is close to the natural frequency of the power system
\cite{ko2026wide,valverde2025forced}. Therefore, reducing only the
overall load variation may not be enough. The frequency content of the
load should also be considered so that components close to vulnerable
grid modes can be specifically reduced.

A hybrid energy storage system (HESS), typically combining a battery
energy storage system (BESS) and a supercapacitor (SC), is a promising
solution for smoothing AI DC power demand. The BESS can handle slower
and longer-lasting power changes, while the SC can respond to fast and
frequent fluctuations, which helps reduce stress on the battery
\cite{powell2026storing,zheng2016hybrid,mohammadi2026grid}. Several recent studies have
investigated coordinated BESS--SC control for this purpose. For example,
authors in \cite{ko2026mitigation} use frequency decomposition
to separate slow and fast load variations and assign them to the BESS
and SC, respectively. This approach is simple to implement, but the
power allocation mainly depends on a predefined frequency separation and
does not explicitly target the load components that are most harmful to
grid oscillations.

To address this issue, author in \cite{chen2026grid} recently proposed a
grid-mode-aware model predictive control (G-MPC) method. The method
includes the vulnerable grid-mode component directly in the control
objective and jointly considers power-envelope limits, ramp-rate limits,
modal suppression, and HESS operating constraints. This allows the HESS
to give higher priority to power components that can strongly excite the
grid. However, the method solves a receding-horizon optimization problem
at every control interval using predicted DC demand. An additional
re-optimization step may also be required to obtain physically consistent
charging and discharging decisions. Thus, although the reported G-MPC
can operate in real time, numerical optimization and load prediction
remain part of the online control process.

To overcome that problem, this paper moves the optimization
from online operation to the offline controller-design stage. A
parameterized grid-mode-aware HESS controller is developed to coordinate
the BESS and SC using load time-scale separation, modal compensation, and
energy-state restoration. Instead of optimizing future BESS and SC power
trajectories at every time step, Bayesian optimization (BO) is used
offline to find suitable controller parameters. Each candidate parameter
set is evaluated through closed-loop simulations over multiple stochastic
AI workload profiles. BO then uses these evaluations to guide the search
toward better parameter values without requiring analytical gradients.
Once the parameters are selected, the same controller is directly
executed online using only filter updates and explicit control equations.
Therefore, no online numerical optimization, re-optimization, or
short-term load forecast is required.

The main contribution of this paper is an offline BO-based
grid-mode-aware HESS control framework for AI DC power smoothing. The
proposed framework keeps explicit awareness of vulnerable grid modes
while moving the computationally intensive parameter search to the
offline stage, leading to a simple online control implementation without
receding-horizon optimization. Numerical studies are used to evaluate
its power-smoothing and grid-frequency performance and to examine the
benefit of explicitly considering modal components in the controller
design.

\vspace{-3mm}
\section{Proposed Methodology}
\label{sec:methodology}

\begin{figure}[t]
\centering
\includegraphics[width=0.9\columnwidth]{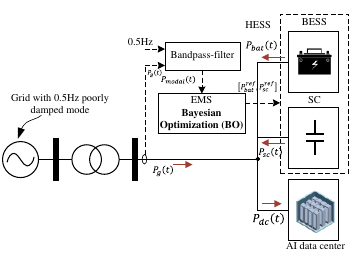}
\caption{Proposed BO-based grid-mode-aware HESS control architecture for AI data-center load smoothing and modal suppression.}
\label{fig_config}
\vspace{-5mm}
\end{figure}

The proposed framework coordinates a BESS and a SC to smooth the highly variable power demand of AI data centers and mitigate load components that may excite poorly damped grid modes, as shown in Fig. \ref{fig_config}. Instead of solving an online receding-horizon optimization problem, a parameterized frequency-decoupling and mode-aware control policy is executed at each sampling instant. Its control parameters are optimized offline using BO, thereby retaining low online computational complexity while accounting for grid-side performance and HESS operating costs.

\subsection{AI Data Center Load and Grid Interaction Modeling}

The total AI data center power demand, $P_{dc}(t)$, consists of the computational workload $P_{work}(t)$, auxiliary demand $P_{aux}$, and an oscillatory component $P_{osc}(t)$:
\begin{equation}
P_{dc}(t)=P_{work}(t)+P_{aux}+P_{osc}(t).
\end{equation}

The computational workload is represented by an Ornstein--Uhlenbeck (OU) mean-reverting stochastic process augmented with compound Poisson burst events \cite{lu2026dynamic}. The utilization state $\eta(t)$ is modeled as
\begin{equation}
\tau_\eta \dot{\eta}
=
-(\eta-\mu_\eta)
+\xi(t)
+\sum_j A_j\delta(t-t_j),
\end{equation}
where $\mu_\eta$ denotes the nominal utilization, $\xi(t)$ represents continuous stochastic variations, and the final term captures burst events occurring at Poisson arrival times $t_j$ with random amplitudes $A_j$. The corresponding workload power is
\begin{equation}
P_{work}(t)
=
p_{base}
+
\eta(t)(p_{full}-p_{base}).
\end{equation}

An oscillatory component
\begin{equation}
P_{osc}(t)=A_{osc}\sin(\omega_{nat}t)
\end{equation}
is superimposed to represent synchronized computational activity capable of exciting a vulnerable grid mode.

The net power observed by the grid is
\begin{equation}
P_g(t)
=
P_{dc}(t)
-
P_{bat}(t)
-
P_{sc}(t),
\label{eq:grid_power}
\end{equation}
where positive storage power denotes discharge toward the data center.

To identify grid-sensitive oscillatory content, $P_g(t)$ is passed through a second-order band-pass filter centered at the vulnerable modal frequency \cite{chen2026grid}:
\begin{align}
\dot{x}_m &= A_m^c x_m+B_m^cP_g,\\
z_m &= C_m^c x_m,
\end{align}
where $x_m$ denotes the filter state and $z_m$ is the extracted modal component. The continuous modal dynamics and converter tracking dynamics are discretized using an exact Zero-Order Hold transformation.

\vspace{-3mm}
\subsection{Parameterized Real-Time HESS Control}

The controller is parameterized by
\begin{equation}
\boldsymbol{\theta}
=
\left[
\tau_{lpf},
\alpha_{sc},
K_{mode},
K_{soc}^{bat},
K_{soc}^{sc}
\right],
\end{equation}
where $\tau_{lpf}$ determines the time-scale separation, $\alpha_{sc}$ controls transient power sharing, $K_{mode}$ scales modal compensation, and $K_{soc}^{bat}$ and $K_{soc}^{sc}$ regulate the storage energy states.

\subsubsection{Frequency Decoupling}

The slowly varying component of the data center load is extracted using a first-order low-pass filter,
\begin{equation}
P_{lpf,k}
=
\rho_fP_{lpf,k-1}
+
(1-\rho_f)P_{dc,k},
\end{equation}
where
\begin{equation}
\rho_f
=
\exp\left(-\frac{\Delta t}{\tau_{lpf}}\right).
\end{equation}

The complementary transient component is
\begin{equation}
P_{tr,k}
=
P_{dc,k}-P_{lpf,k}.
\end{equation}

This decomposition enables fast load variations to be preferentially assigned to the SC while slower energy-balancing requirements are handled primarily by the BESS.

\subsubsection{Energy-State Restoration and Power Allocation}

To prevent gradual drift of the storage energy states, proportional restoration terms are defined as
\begin{align}
P_{soc,k}^{bat}
&=
K_{soc}^{bat}
(E_{bat,0}-E_{bat,k}),\\
P_{soc,k}^{sc}
&=
K_{soc}^{sc}
(E_{sc,0}-E_{sc,k}).
\end{align}

The SC power reference is
\begin{equation}
P_{sc,k}^{ref}
=
\alpha_{sc}P_{tr,k}
+
K_{mode}z_{m,k}
+
P_{soc,k}^{sc},
\end{equation}
where the first term assigns a fraction of the transient component to the SC and the second introduces mode-aware compensation.

The BESS reference is
\begin{equation}
P_{bat,k}^{ref}
=
(1-\alpha_{sc})P_{tr,k}
+
(P_{lpf,k}-P_{center})
+
P_{soc,k}^{bat},
\end{equation}
where $P_{center}$ denotes the desired center of the grid-side operating range. Thus, the BESS absorbs the remaining transient component while also correcting slower deviations of the grid demand from its reference level.

The power references are first restricted by the rated converter limits. The finite converter response is represented using first-order tracking dynamics,
\begin{equation}
\widetilde P_{i,k}^{act}
=
\rho_iP_{i,k-1}^{act}
+
(1-\rho_i)P_{i,k}^{ref},
\qquad
i\in\{bat,sc\},
\end{equation}
with
\begin{equation}
\rho_i
=
\exp\left(-\frac{\Delta t}{\tau_i}\right).
\end{equation}

\subsubsection{Energy-Feasible Storage Operation}

In addition to the rated converter limits, the commanded charge and discharge powers are constrained by the energy currently available in each storage device. Let
\begin{equation}
\Delta t_h=\frac{\Delta t}{3600}
\end{equation}
denote the sampling interval expressed in hours. The energy-dependent discharge and charging limits are
\begin{align}
P_{i,k}^{dis,E}
&=
\frac{(E_{i,k}-\underline E_i)\eta_i}{\Delta t_h},\\
P_{i,k}^{ch,E}
&=
\frac{\overline E_i-E_{i,k}}
{\eta_i\Delta t_h}.
\end{align}

The feasible instantaneous limits are therefore
\begin{align}
\overline P_{i,k}^{dis}
&=
\min(\overline P_i,P_{i,k}^{dis,E}),\\
\overline P_{i,k}^{ch}
&=
\min(\overline P_i,P_{i,k}^{ch,E}),
\end{align}
and the actual power is obtained by projecting the converter output onto the feasible interval:
\begin{equation}
P_{i,k}^{act}
=
\Pi_{
[-\overline P_{i,k}^{ch},
\overline P_{i,k}^{dis}]
}
(\widetilde P_{i,k}^{act}).
\end{equation}

This prevents either storage device from delivering power below its minimum energy level or absorbing power above its maximum energy level.

The actual power is separated into nonnegative charging and discharging components,
\begin{align}
P_{i,k}^{dis}
&=
\max(0,P_{i,k}^{act}),\\
P_{i,k}^{ch}
&=
\max(0,-P_{i,k}^{act}),
\end{align}
which inherently prevents simultaneous charging and discharging.

The energy dynamics are then
\begin{equation}
E_{i,k+1}
=
E_{i,k}
+
\Delta t_h
\left(
\eta_iP_{i,k}^{ch}
-
\frac{1}{\eta_i}P_{i,k}^{dis}
\right),
\end{equation}
subject to
\begin{equation}
\underline E_i
\leq E_{i,k}\leq\overline E_i.
\end{equation}

\subsection{Grid and HESS Performance Objective}

For a given parameter vector $\boldsymbol{\theta}$ and workload realization, the controller is evaluated using a cumulative objective combining grid-side performance and HESS operating costs:
\begin{equation}
J(\boldsymbol{\theta})
=
\sum_{k=1}^{N}
\Delta t_h
\left(
J_{grid,k}
+
J_{hess,k}
\right).
\end{equation}

The grid-side term penalizes violations of the desired power envelope, grid ramp-rate limit, and modal-power limit:
\begin{equation}
J_{grid,k}
=
c_p\epsilon_{p,k}^2
+
c_r\epsilon_{r,k}^2
+
c_m\epsilon_{m,k}^2,
\end{equation}
where
\begin{equation}
\epsilon_{p,k}
=
\max
\left(
0,
\underline P_g-P_{g,k},
P_{g,k}-\overline P_g
\right),
\end{equation}
\begin{equation}
\epsilon_{r,k}
=
\max
\left(
0,
\left|
\frac{P_{g,k}-P_{g,k-1}}{\Delta t}
\right|
-\overline R_g
\right),
\end{equation}
and
\begin{equation}
\epsilon_{m,k}
=
\max
\left(
0,
|z_{m,k}|-\overline z_m
\right).
\end{equation}

These quantities are treated as soft grid-side requirements. Consequently, temporary violations remain feasible but are penalized according to their magnitude.

The HESS-side cost consists of degradation and power-ramping terms:
\begin{equation}
J_{hess,k}
=
J_{hess,k}^{deg}
+
J_{hess,k}^{ramp}.
\end{equation}

The degradation term penalizes storage energy throughput,
\begin{equation}
\begin{split}
J_{hess,k}^{deg}
={}&
c_{bat}^{deg}
\left(
\eta_{bat}P_{bat,k}^{ch}
+
\frac{P_{bat,k}^{dis}}{\eta_{bat}}
\right)
\\
&+
c_{sc}^{deg}
\left(
\eta_{sc}P_{sc,k}^{ch}
+
\frac{P_{sc,k}^{dis}}{\eta_{sc}}
\right),
\end{split}
\end{equation}
while the ramping term is
\begin{equation}
\begin{split}
J_{hess,k}^{ramp}
={}&
c_{bat}^{\Delta}
\left(
\frac{P_{bat,k}^{act}-P_{bat,k-1}^{act}}
{\Delta t}
\right)^2
\\
&+
c_{sc}^{\Delta}
\left(
\frac{P_{sc,k}^{act}-P_{sc,k-1}^{act}}
{\Delta t}
\right)^2.
\end{split}
\end{equation}

The combination of these terms encourages grid-compatible power trajectories while limiting unnecessary storage throughput and rapid cycling.

\subsection{Offline Tuning via Bayesian Optimization}

The relationship between $\boldsymbol{\theta}$ and the cumulative closed-loop cost is nonlinear because of the stochastic workload, modal dynamics, converter response, saturation limits, energy constraints, and nonlinear penalty terms. The controller tuning problem is therefore treated as a bounded black-box optimization problem:
\begin{equation}
\boldsymbol{\theta}^{*}
=
\arg\min_{\boldsymbol{\theta}\in\Theta}
\mathbb{E}
\left[
J(\boldsymbol{\theta};P_{dc})
\right].
\end{equation}

To approximate the expectation over stochastic AI workloads, each candidate controller is evaluated on a fixed collection of independently generated training profiles. The BO objective is the corresponding sample-average cost:
\begin{equation}
\widehat J_{train}(\boldsymbol{\theta})
=
\frac{1}{N_s}
\sum_{s=1}^{N_s}
J
\left(
\boldsymbol{\theta};
P_{dc}^{(s)}
\right).
\label{eq:bo_sample_average}
\end{equation}

Using the same workload realizations for all candidate controllers provides a consistent comparison between parameter sets and avoids introducing additional Monte-Carlo variation during the BO iterations.

A Gaussian Process (GP) is used as a probabilistic surrogate model of the objective landscape. The controller parameters are normalized before fitting the GP, while the observed objective values are standardized to improve numerical conditioning. Since the implementation maximizes the negative objective, the surrogate is defined for
\begin{equation}
y(\boldsymbol{\theta})
=
-\widehat J_{train}(\boldsymbol{\theta}),
\end{equation}
with
\begin{equation}
y(\boldsymbol{\theta})
\sim
\mathcal{GP}
\left(
m(\boldsymbol{\theta}),
k(\boldsymbol{\theta},\boldsymbol{\theta}')
\right).
\end{equation}

Following an initial set of controller evaluations, the logarithmic Expected Improvement (LogEI) acquisition function is used to identify the next parameter vector:
\begin{equation}
\boldsymbol{\theta}_{next}
=
\arg\max_{\boldsymbol{\theta}\in\Theta}
a_{\mathrm{LogEI}}(\boldsymbol{\theta}).
\end{equation}

After each new evaluation, the observation is appended to the training data and the GP surrogate is refitted. The final controller corresponds to the evaluated parameter vector achieving the lowest sample-average training cost.

To assess generalization, the selected controller is subsequently evaluated on an independent set of stochastic workload realizations that are excluded from the BO tuning process. The detailed BO configuration, parameter ranges, number of training and test profiles, and numerical performance are reported in Section~\ref{sec:numerical_analysis}.

\section{Numerical Analysis}
\label{sec:numerical_analysis}

This section evaluates the proposed coordinated control framework for AI data center power smoothing and mitigation of grid-sensitive oscillatory load components. The controller parameters are tuned using BO over multiple stochastic AI workload realizations, and the resulting policies are subsequently evaluated on unseen load profiles. The numerical analysis focuses on grid-side power shaping, BO convergence, frequency response, and attenuation of the dominant oscillatory load component.

\subsection{Simulation Setup and Parameter Settings}

\subsubsection{Aggregate Power Grid and AI Data Center Load Model}

To evaluate the impact of highly variable AI workloads on grid frequency stability, the data center is connected to an aggregate power system represented by linearized swing and governor dynamics. The grid parameters are selected such that the resulting system contains a poorly damped natural mode at
$f_{nat}=0.5$~Hz.

The AI data center demand profile, $P_{dc}(t)$, is generated using an Ornstein--Uhlenbeck (OU) stochastic process combined with compound Poisson burst events \cite{lu2026dynamic}. The OU component represents continuously varying computational utilization, whereas the Poisson component introduces intermittent workload surges. In addition, a periodic component with an amplitude of $20$~MW at $0.5$~Hz is superimposed on the stochastic demand to explicitly excite the vulnerable grid mode. Each simulation covers $300$~s with a sampling interval of $\Delta t=0.25$~s.

\subsubsection{HESS Configuration and Parameters}

The Hybrid Energy Storage System (HESS) consists of a BESS with a power limit of $40$~MW and an operational energy range of $4$--$36$~MWh, together with a supercapacitor (SC) rated at $20$~MW with an operational energy range of $0.03$--$0.27$~MWh. Their initial stored energies are $20$~MWh and $0.15$~MWh, respectively. The BESS has a converter response time constant of $\tau_{bat}=0.08$~s and the SC has a faster response of $\tau_{sc}=0.05$~s.

The storage dynamics account for charging/discharging efficiencies and enforce both rated-power limits and instantaneous power limits imposed by the available stored energy. Therefore, neither storage device can discharge below its minimum energy level or charge beyond its maximum energy level. The principal system parameters and objective coefficients are summarized in Table~\ref{tab:parameters} \cite{chen2026grid}.

\begin{table}[htbp]
\centering
\scriptsize
\caption{System Parameters and Cost Coefficients}
\label{tab:parameters}
\begin{tabular}{@{}lll@{}}
\toprule
\textbf{Category} & \textbf{Parameter} & \textbf{Value} \\ 
\midrule

\multirow{4}{*}{\textbf{Grid \& Load}} 
 & Base Load ($P_{\mathrm{base}}$) & $25.0$ MW \\
 & Grid Power Envelope & $[120.0,130.0]$ MW \\
 & Ramp Rate Limit ($R_{g,\max}$) & $1.0$ MW/s \\
 & Grid Natural Freq. ($f_{\mathrm{nat}}$) & $0.5$ Hz \\ 
\midrule

\multirow{5}{*}{\textbf{BESS}} 
 & Power Limit ($\overline{P}_{\mathrm{bat}}$) & $40.0$ MW \\
 & Energy Limits ($\underline{E}_{\mathrm{bat}},\overline{E}_{\mathrm{bat}}$) 
 & $4.0,36.0$ MWh \\
 & Efficiency ($\eta_{\mathrm{bat}}$), Response ($\tau_{\mathrm{bat}}$) 
 & $0.97,0.08$ s \\
 & Degradation Cost ($c_{\mathrm{bat}}^{\mathrm{deg}}$) 
 & \$20.0/MWh \\
 & Ramping Penalty ($c_{\mathrm{bat}}^{\mathrm{ramp}}$) 
 & \$2.0/$(\mathrm{MW/s})^2\mathrm{h}$ \\ 
\midrule

\multirow{5}{*}{\textbf{SC}} 
 & Power Limit ($\overline{P}_{\mathrm{sc}}$) & $20.0$ MW \\
 & Energy Limits ($\underline{E}_{\mathrm{sc}},\overline{E}_{\mathrm{sc}}$) 
 & $0.03,0.27$ MWh \\
 & Efficiency ($\eta_{\mathrm{sc}}$), Response ($\tau_{\mathrm{sc}}$) 
 & $0.98,0.05$ s \\
 & Degradation Cost ($c_{\mathrm{sc}}^{\mathrm{deg}}$) 
 & \$5.0/MWh \\
 & Ramping Penalty ($c_{\mathrm{sc}}^{\mathrm{ramp}}$) 
 & \$0.2/$(\mathrm{MW/s})^2\mathrm{h}$ \\ 
\midrule

\multirow{3}{*}{\textbf{Obj. Weights}} 
 & Env. Violation Penalty ($c_p$) 
 & \$10.0/$\mathrm{MW}^2\mathrm{h}$ \\
 & Ramp Violation Penalty ($c_r$) 
 & \$3.0/$(\mathrm{MW/s})^2\mathrm{h}$ \\
 & Modal Power Penalty ($c_m$) 
 & \$100.0/$\mathrm{MW}^2\mathrm{h}$ \\ 
\bottomrule

\end{tabular}
\vspace{-3mm}
\end{table}

\subsubsection{Bayesian Optimization and Evaluation Protocol}

The controller contains five tunable parameters,
\[
\boldsymbol{\theta}
=
[\tau_{\mathrm{LPF}},
\alpha_{sc},
K_{\mathrm{mode}},
K_{\mathrm{soc,bat}},
K_{\mathrm{soc,sc}}],
\]
with corresponding search intervals
\[
\tau_{\mathrm{LPF}}\in[0.5,20], \qquad
\alpha_{sc}\in[0,1],
\]
\[
K_{\mathrm{mode}}\in[0,10], \qquad
K_{\mathrm{soc,bat}}\in[0,10], \qquad
K_{\mathrm{soc,sc}}\in[0,100].
\]

Rather than optimizing the controller for a single stochastic realization, the BO objective is defined as the mean control cost over $10$ fixed training workload profiles,
\[
J_{\mathrm{train}}(\boldsymbol{\theta})
=
\frac{1}{10}
\sum_{s=1}^{10}
J\left(\boldsymbol{\theta},P_{dc}^{(s)}\right).
\]
This formulation reduces overfitting to a particular workload realization. A Gaussian-process surrogate with normalized inputs and standardized outputs is employed together with the logarithmic Expected Improvement acquisition function. The optimization begins with $20$ randomly sampled initial designs, followed by $50$ sequential BO evaluations, resulting in $70$ objective-function evaluations per case.

After optimization, the selected parameters are evaluated on $20$ independent workload realizations that are not used during training. The time-domain and frequency-domain results presented below correspond to the first unseen test realization.

To assess the contribution of the individual objective components, three cases are considered:
\begin{itemize}
    \item \textbf{Case 1 (Complete Model):} power-envelope, grid-ramp, and modal penalties are all included in the optimization objective.
    \item \textbf{Case 2 (Mode Suppression Only):} the explicit power-envelope and grid-ramp penalties are disabled, while the $0.5$~Hz modal penalty is retained.
    \item \textbf{Case 3 (No Modal Penalty):} the power-envelope and grid-ramp penalties are retained, while the explicit modal penalty is disabled.
\end{itemize}

It should be noted that the controller structure itself retains the low-pass power-sharing mechanism in all three cases. Consequently, Cases 2 and 3 still exhibit a degree of general power smoothing even when the corresponding objective terms are disabled.

\subsection{Performance of the Proposed Coordinated Control}

\begin{figure}[htbp]
    \centering
    \includegraphics[width=\columnwidth]{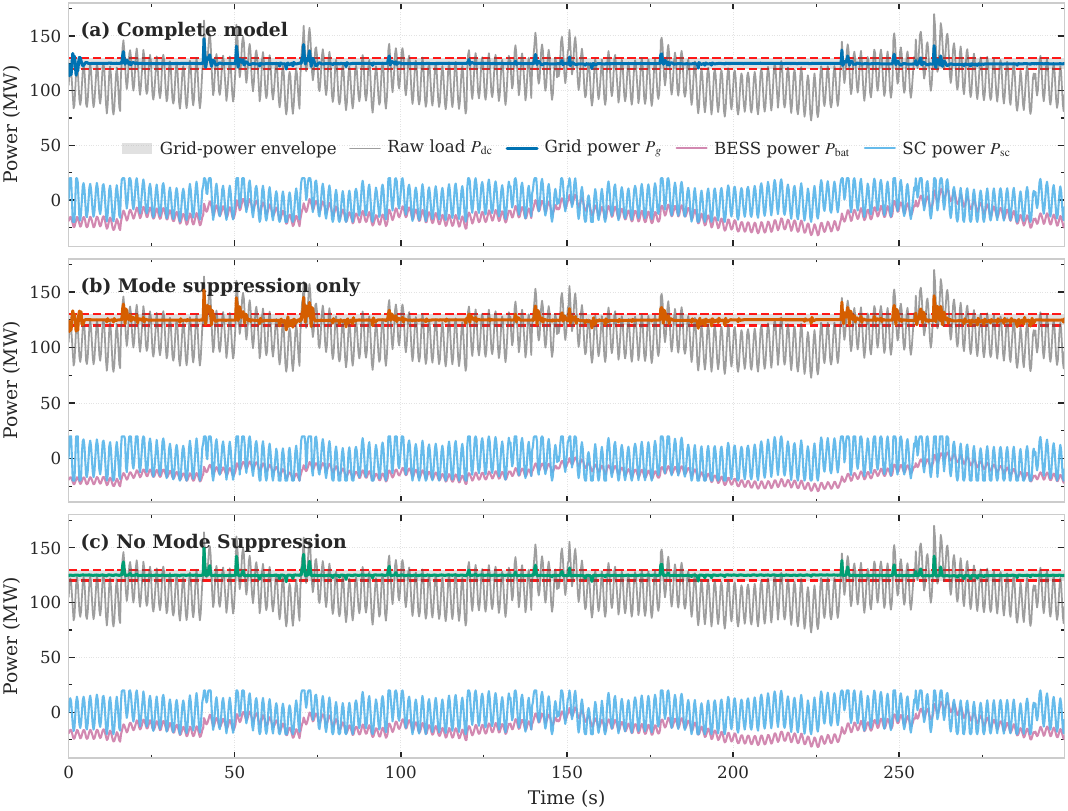}
    \caption{Power-smoothing performance on an unseen AI workload realization. Raw data center demand $P_{dc}(t)$, resulting grid-side power $P_g(t)$, BESS power $P_{bat}(t)$, and SC power $P_{sc}(t)$ are shown for Cases~1--3. Dashed horizontal lines indicate the $120$--$130$~MW desired grid-power envelope.}
    \label{fig:power_smoothing}
    \vspace{-5mm}
\end{figure}

Fig.~\ref{fig:power_smoothing} shows the raw AI data center demand, the resulting grid-side power, and the BESS and SC power trajectories for the three cases. The considered unseen test realization contains pronounced stochastic variations superimposed on the $0.5$~Hz periodic component, with the raw demand varying approximately between $75$~MW and $160$~MW.

For Case~1, the HESS substantially reshapes the raw demand and maintains $P_g$ close to the desired $120$--$130$~MW operating region for most of the simulation. Only short-duration excursions occur during the strongest workload bursts. Case~3 exhibits similarly strong low-frequency power regulation because the power-envelope and ramp penalties remain active. Case~2 also keeps the grid-side power centered near the desired operating region, despite the removal of the explicit envelope penalty. This behavior results from the underlying controller structure, which includes the low-pass reference term $(p_{\mathrm{LPF}}-125)$ and therefore inherently regulates the grid-side power around $125$~MW. However, Case~2 exhibits more noticeable transient excursions than Case~1.

The complementary operation of the two storage technologies is also visible in Fig.~\ref{fig:power_smoothing}. The SC responds predominantly to the fast periodic component, repeatedly charging and discharging at high frequency, whereas the BESS provides the slower energy-balancing component associated with sustained deviations of the workload from the grid-side reference. This separation reduces the exposure of the BESS to the fastest power fluctuations while allowing the SC to absorb the high-frequency component.

\begin{figure}[htbp]
    \centering
    \includegraphics[width=\columnwidth]{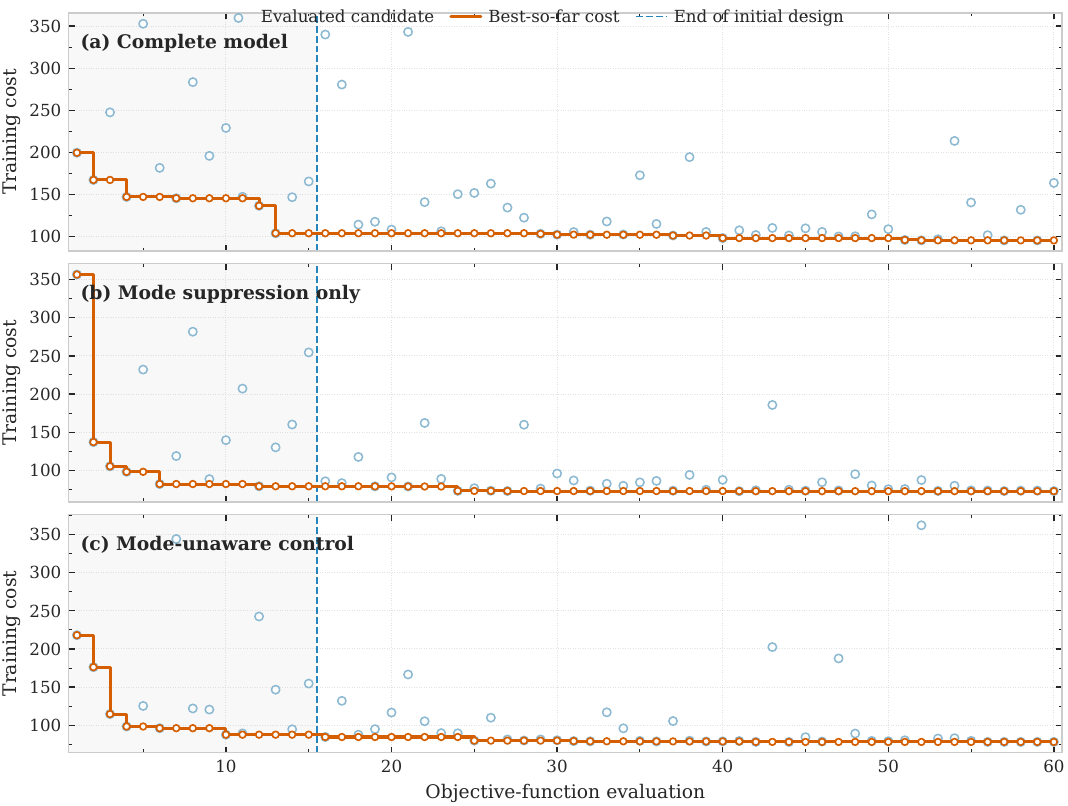}
    \caption{Bayesian Optimization convergence for Cases~1--3. The light curve shows the mean training cost of each evaluated controller, while the solid curve shows the best-so-far cost. The vertical dashed line marks the end of the $20$-sample initial design and the beginning of the sequential BO stage.}
    \label{fig:bo_convergence}
    \vspace{-5mm}
\end{figure}

Fig.~\ref{fig:bo_convergence} presents the BO convergence behavior for the three cases. The first $15$ evaluations correspond to the initial randomly sampled controller designs, after which the Gaussian-process surrogate and acquisition function are used to select new candidates. The individual candidate costs exhibit substantial variability, reflecting the nonlinear interaction between the five controller parameters. In contrast, the best-so-far objective decreases monotonically as progressively improved controller settings are discovered.

For Case~1, the best mean training cost decreases rapidly during the initial design stage from above $200$ to approximately $100$, followed by a smaller additional improvement during the BO stage to approximately $90$. Case~2 converges from an initial cost exceeding $350$ to approximately $50$, while Case~3 reaches a similar final value after a pronounced improvement shortly after the BO stage begins. These results indicate that the majority of the improvement is obtained within a relatively small number of evaluations, while the later BO iterations primarily refine the solution.

The absolute objective values of the three cases should not be directly interpreted as a ranking of overall control quality because different penalty components are active in each case. The convergence curves are instead used to demonstrate that BO successfully identifies progressively improved parameter sets within each respective objective formulation.

\subsection{Grid Frequency Response and Oscillation Mitigation}

\begin{figure}[htbp]
    \centering
    \includegraphics[width=\columnwidth]{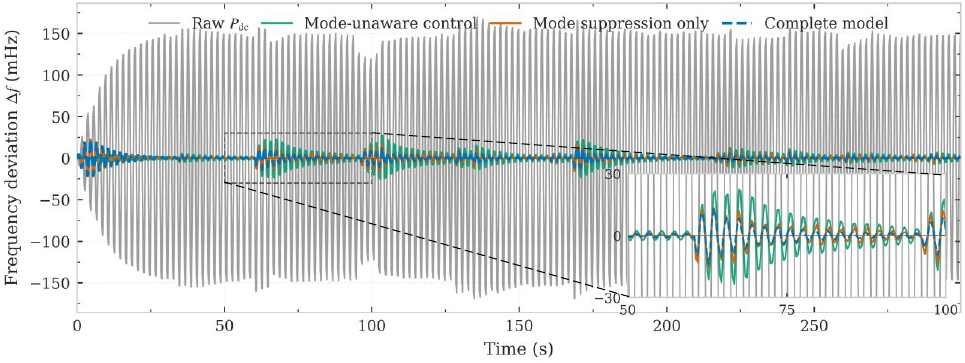}
    \caption{Grid-frequency response for the raw AI data center demand and for the HESS-filtered grid-side power obtained under Cases~1--3 on the unseen test realization.}
    \label{fig:grid_frequency}
    \vspace{-3mm}
\end{figure}

The resulting grid-frequency trajectories are shown in Fig.~\ref{fig:grid_frequency}. Without HESS compensation, the periodic $0.5$~Hz component of the AI data center demand continuously excites the poorly damped grid mode, producing sustained oscillations of approximately $59.85$--$60.15$~Hz after the initial transient.

All three optimized controllers significantly reduce these oscillations. Cases~1 and 2 maintain the grid frequency very close to the nominal $60$~Hz value, with only short transient deviations associated with stochastic workload bursts. Case~3 also provides a substantial reduction relative to the raw demand, but more visible residual oscillations remain during several intervals. The result demonstrates that generic power smoothing alone already provides substantial frequency support, while explicit consideration of the vulnerable mode further improves attenuation of the resonant component.

\begin{figure}[htbp]
    \centering
    \includegraphics[width=\columnwidth]{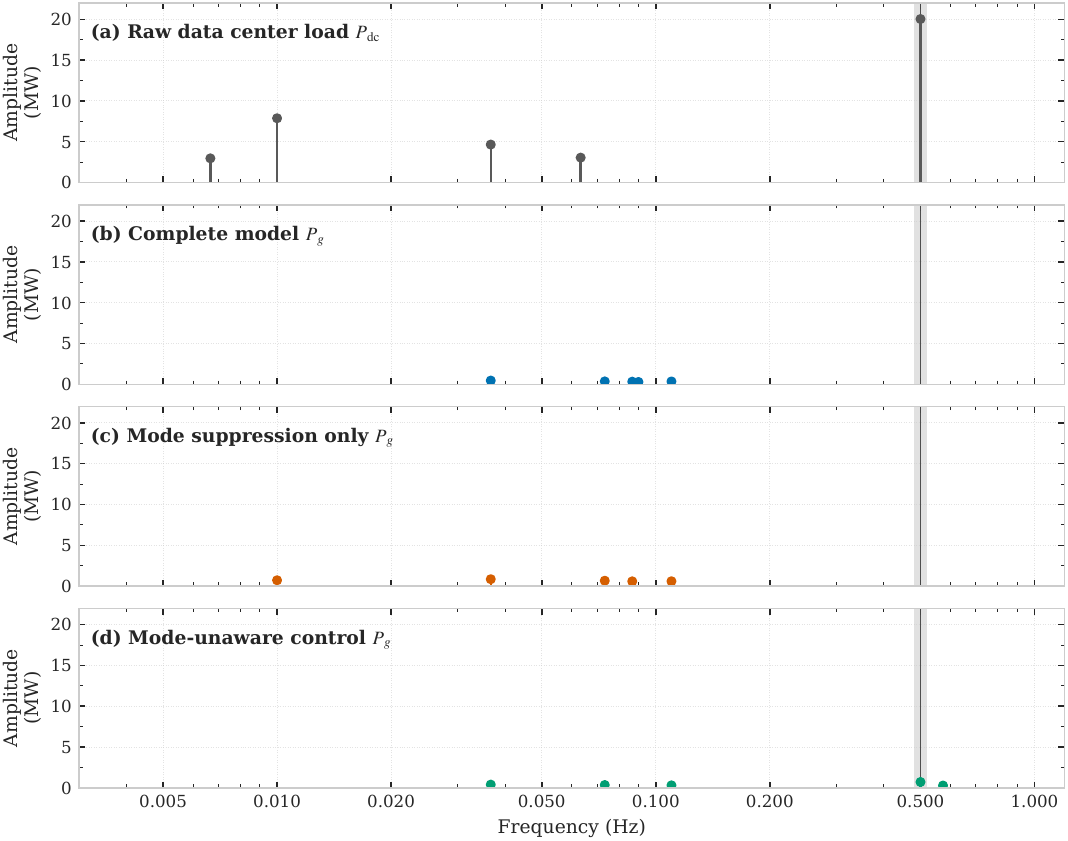}
    \caption{FFT comparison of the five dominant non-DC frequency components of the raw data center demand and the resulting grid-side power for Cases~1--3. The shaded region highlights the vulnerable grid-mode neighborhood around $0.5$~Hz.}
    \label{fig:fft_analysis}
    \vspace{-5mm}
\end{figure}

The frequency-domain behavior is examined further in Fig.~\ref{fig:fft_analysis}, which shows the five dominant non-DC spectral components of the raw and controlled power signals. The raw data center demand contains a dominant $20$~MW component at precisely $0.5$~Hz, corresponding to the imposed periodic computational-load oscillation. This component is considerably larger than the remaining stochastic spectral components and therefore dominates the excitation of the grid mode.

The HESS strongly attenuates this component in all three cases. In Case~1 and Case~2, the $0.5$~Hz component no longer appears among the five dominant spectral components, indicating particularly strong suppression of the targeted mode. The remaining dominant components are primarily located at lower frequencies and remain below approximately $1$~MW. Case~3 also achieves substantial attenuation because the low-pass power-sharing controller removes a large fraction of the fast load variation. However, a residual component remains visible near $0.5$~Hz.

These results refine the interpretation of the ablation study. Explicit modal awareness is not solely responsible for all oscillation attenuation, since the HESS power-smoothing structure itself already suppresses a significant portion of the resonant variation. Nevertheless, including the modal penalty further reduces the spectral content around the vulnerable grid mode. Case~1 therefore provides the most complete formulation by jointly considering grid-power regulation, ramp-rate performance, storage operation, and mode-sensitive oscillation mitigation.

\vspace{-3mm}
\section{Conclusion}

This study demonstrates that a coordinated HESS effectively mitigates extreme AI data center power volatility. By separating high-frequency transients from low-frequency demand shifts, the control policy leverages supercapacitors and batteries to optimize performance and minimize storage degradation. Crucially, numerical analysis proves explicit targeted mode suppression is more vital for grid frequency stability than generic load smoothing. The optimized framework strictly regulates raw data center demand primarily within the 120-130 MW operating region, allowing only brief deviations during severe workload bursts. Targeting the vulnerable 0.5 Hz grid mode stabilizes sustained deviations and maintains frequency near the nominal 60 Hz. Furthermore, offline BO effectively navigates stochastic AI workload uncertainty, showing strong convergence by reducing mean training costs. This offers a robust, computationally efficient, and forecasting-independent solution for grid-compliant operation without real-time prediction.

\vspace{-5mm}
\bibliographystyle{IEEEtran}
\bibliography{Ref}

\end{document}